\documentclass[acmlarge]{acmart}

\usepackage{graphicx}
\usepackage{float}
\usepackage{longtable}
\usepackage{tabularx}
\usepackage{xcolor}
\usepackage{caption}
\usepackage{subcaption}
\usepackage{gensymb}
\usepackage{diagbox}

\AtBeginDocument{%
  }

\setcopyright{acmlicensed}
\copyrightyear{2024}
\acmYear{2024}
\acmDOI{XXXXXXX.XXXXXXX}

\acmJournal{JATS}
\acmVolume{37}
\acmNumber{4}
\acmArticle{111}
\acmMonth{11}

\begin{document}

%%
%% The "title" command has an optional parameter,
%% allowing the author to define a "short title" to be used in page headers.
\title{Enhancing Disaster Resilience with UAV-Assisted Edge Computing: A Reinforcement Learning Approach to Managing Heterogeneous Edge Devices}

%%
%% The "author" command and its associated commands are used to define
%% the authors and their affiliations.
%% Of note is the shared affiliation of the first two authors, and the
%% "authornote" and "authornotemark" commands
%% used to denote shared contribution to the research.
\author{Talha Azfar}
%\authornote{Both authors contributed equally to this research.}
\email{azfart@rpi.edu}
\orcid{0000-0002-1293-5036}
\author{Kaicong Huang}
%\orcid{0000-0001-1234-5678}
%\authornotemark[1]
\email{huangk10@rpi.edu}
\author{Ruimin Ke}
%\authornotemark[1]
\email{ker@rpi.edu}

\affiliation{%
  \institution{Rensselaer Polytechnic Institute}
  \city{Troy}
  \state{New York}
  \country{USA}
}

% \author{Lars Th{\o}rv{\"a}ld}
% \affiliation{%
%   \institution{The Th{\o}rv{\"a}ld Group}
%   \city{Hekla}
%   \country{Iceland}}
% \email{larst@affiliation.org}

% \author{Valerie B\'eranger}
% \affiliation{%
%   \institution{Inria Paris-Rocquencourt}
%   \city{Rocquencourt}
%   \country{France}
% }

% \author{Aparna Patel}
% \affiliation{%
%  \institution{Rajiv Gandhi University}
%  \city{Doimukh}
%  \state{Arunachal Pradesh}
%  \country{India}}

% \author{Huifen Chan}
% \affiliation{%
%   \institution{Tsinghua University}
%   \city{Haidian Qu}
%   \state{Beijing Shi}
%   \country{China}}

% \author{Charles Palmer}
% \affiliation{%
%   \institution{Palmer Research Laboratories}
%   \city{San Antonio}
%   \state{Texas}
%   \country{USA}}
% \email{cpalmer@prl.com}

% \author{John Smith}
% \affiliation{%
%   \institution{The Th{\o}rv{\"a}ld Group}
%   \city{Hekla}
%   \country{Iceland}}
% \email{jsmith@affiliation.org}

% \author{Julius P. Kumquat}
% \affiliation{%
%   \institution{The Kumquat Consortium}
%   \city{New York}
%   \country{USA}}
% \email{jpkumquat@consortium.net}

%%
%% By default, the full list of authors will be used in the page
%% headers. Often, this list is too long, and will overlap
%% other information printed in the page headers. This command allows
%% the author to define a more concise list
%% of authors' names for this purpose.
%\renewcommand{\shortauthors}{Azfar et al.}

%%
%% The abstract is a short summary of the work to be presented in the
%% article.
\begin{abstract}
Edge sensing and computing is rapidly becoming part of intelligent infrastructure architecture leading to operational reliance on such systems in disaster or emergency situations. In such scenarios there is a high chance of power supply failure due to power grid issues, and communication system issues due to base stations losing power or being damaged by the elements, e.g., flooding, wildfires etc. Mobile edge computing in the form of unmanned aerial vehicles (UAVs) has been proposed to provide computation offloading from these devices to conserve their battery, while the use of UAVs as relay network nodes has also been investigated previously. This paper considers the use of UAVs with further constraints on power and connectivity to prolong the life of the network while also ensuring that the data is received from the edge nodes in a timely manner. Reinforcement learning is used to investigate numerous scenarios of various levels of power and communication failure. This approach is able to identify the device most likely to fail in a given scenario, thus providing priority guidance for maintenance personnel. The evacuations of a rural town and urban downtown area are also simulated to demonstrate the effectiveness of the approach at extending the life of the most critical edge devices.
\end{abstract}

%%
%% The code below is generated by the tool at http://dl.acm.org/ccs.cfm.
%% Please copy and paste the code instead of the example below.
%%


%%
%% Keywords. The author(s) should pick words that accurately describe
%% the work being presented. Separate the keywords with commas.

\received{29 November 2024}
% \received[revised]{12 March 2009}
% \received[accepted]{5 June 2009}

%%
%% This command processes the author and affiliation and title
%% information and builds the first part of the formatted document.
\maketitle

\renewcommand{\arraystretch}{1.25}
\section{Introduction}

In the face of natural disasters, remote locations, or emergency scenarios, the availability of reliable infrastructure remains a critical challenge. Traditional infrastructure systems often falter in such situations due to power outages, communication breakdowns, and the inability to conduct timely maintenance. In these contexts, edge infrastructure devices play a pivotal role in maintaining connectivity, collecting vital data, and facilitating emergency response efforts in urban areas \cite{higashino2017edge} and remote hostile environments \cite{hussain2019federated}. 

Edge computing devices usually perform crucial tasks involving sensing, processing, and communication while consuming relatively low power. This could be as simple as recording and updating the temperature, or as complex as real-time multi-vehicle detection and tracking from surveillance cameras. These edge devices can be instrumental in disaster response by monitoring the situation, detecting urgent areas of concern, and coordinating the emergency response \cite{aboualola2023edge}.
Some challenges faced by these edge devices in remote locations are sporadic or otherwise limited power supply and communication channels. Similarly, in disaster scenarios such as floods or wildfires, power grids and wireless communication base stations may partly or completely fail, leaving the edge devices to run on backup power and wait for alternative communication channels such as mobile base stations until the usual systems are restored. 

However, getting to the locations of the devices and providing the necessary services may not be feasible in the short term, especially in emergency situations where other tasks may take priority. Recent works have proposed an innovative solution that tackles both the power consumption and communication constraints in the form of an unmanned aerial vehicle (UAV) mobile edge computing system \cite{wang2021deep}, in which a UAV flies within communication range of the edge devices and offloads their computation tasks thus saving their battery, and simultaneously acts as a high gain relay node to the base station which is not reachable by the low power edge devices. The UAV is not bound by the same limitations as the devices in the field since it can return to its base for charging at any time. 

While UAV relay nodes as part of the communication network and UAV mobile edge computing nodes have been considered separately, the idea of data criticality in emergency situations needs more attention. During a disaster recovery process, intelligent edge infrastructure may provide crucial data to responders, and is only useful if the communication is up to date. Data criticality then emerges as a pivotal concern in these networks, reflecting the diverse and important nature of data handled. If power and communications are both affected by the disaster, then there is no guarantee that data is being received from all edge nodes on time. This motivates the construction of a strategy that pays special attention to the age of the uncommunicated data held in the edge devices that have lost communication. Together with the power loss constraint, it becomes an interesting problem to study with reinforcement learning to discover the best strategy in any given configuration of devices. Another important aspect missing from the existing literature is the simulation of the direct impacts of natural disasters and the associated evacuation of individuals utilizing road networks. Furthermore, it raises the critical question of how to prioritize edge devices based on their utility and effectiveness in such scenarios.

In this paper we explore the feasibility and efficacy of integrating UAVs and reinforcement learning (RL) techniques into the deployment and management of edge infrastructure devices in rural areas and emergency scenarios by varying the possible constraints and configurations to reflect real-world challenges. Specifically, we address the issue of losing power and communication in disaster stricken areas and show that the UAV can learn to extend the life of the devices. First, the problem is mathematically formulated with physical constraints on the device energy and data age. Then, we use reinforcement learning with different reward functions to identify the most general method of solving the problem. 
We also simulate the evacuation of a rural town and an urban downtown area, and study how the UAV can prioritize edge devices based on their proximity to the traffic flow. The contributions of this paper are summarized as follows:
\begin{itemize}
    \item Handling Heterogeneous Edge Devices: The paper addresses the challenge of managing a diverse set of edge devices with varying computational capabilities, power constraints, and communication abilities proposing a UAV-assisted system to uniformly manage these differences while maintaining network resilience.
    
    \item Mitigating Power and Communication Outages: By utilizing UAVs to provide both computational offloading and communication relays, the study offers a solution to the frequent power and connection outages experienced by edge devices during disasters, ensuring continuous data collection and connectivity.
    
    \item Reinforcement Learning for Strategic Optimization: The system employs deep Q-network reinforcement learning to develop strategies for optimizing UAV operations, including prioritizing critical data delivery and extending the operational life of the network in various disaster scenarios.

    \item Scenario-Based Evaluation: The study conducts extensive simulations, including realistic disaster evacuation scenarios, to demonstrate the effectiveness of the proposed approach in prioritizing critical data delivery and maintaining infrastructure functionality under diverse conditions of power and communication availability.
\end{itemize}

\section{Literature Review}

Researchers have long explored the integration of UAVs with cellular network infrastructure to optimize energy efficiency while ensuring the maintenance of quality of service \cite{naqvi2018drone}\cite{tran2022throughput}, while the mixed architecture of terrestrial and aerial connectivity has featured in numerous studies about infrastructure resilience in the face of disasters for preparedness, assessment, and response and recovery \cite{erdelj2017help}. 
UAVs have been considered as relay nodes in communication networks to provide greater, more resilient coverage, for example in \cite{zeng2016throughput}, where an iterative convex optimization algorithm finds the best power allocation and trajectory which also maximizes network throughput. The use of UAVs as an intermediate cloudlet layer in communication or distributed processing networks has been discussed in many architectures precisely for infrastructure resilience during natural or man-made disasters \cite{kaleem2019uav}. Communication disruptions in disaster scenarios are addressed in \cite{xu2020big} with LoRaWAN equipped UAV mobile edge computing which significantly increases channel capacity by modeling the air-to-ground and remote-to-air connections individually to formulate a Markov chain for task assignment and queue management. 

There has been significant research focused on the exact dynamics of UAV mobile edge computing to minimize energy consumption by optimizing the trajectory of the UAVs and computing resource allocations. The energy efficiency was maximized by jointly optimizing UAV trajectory, transmission power, and load allocation in \cite{li2020energy} using successive convex approximations on multiple subproblems. Convex optimizations and reinforcement learning have been used by \cite{wang2021deep} which employ block coordinate descent to solve the convex problem for a given scenario, and then use those insights in a deep Q network reinforcement learning framework for generalized rapid decision making. Similarly, \cite{sun2023optimal} formulate a time slot based Markov decision process to model the problem and then use a double deep Q-learning algorithm to minimize energy consumption and communication overhead, which maximizes the lifetime of the system before device failure. A two layer optimization problem was constructed for task scheduling and trajectory optimization for a multi-UAV mobile edge computing platform in \cite{luo2021optimization}.

The vehicle path is optimized for maximum data collection from wireless sensor networks with energy harvesting nodes in \cite{mehrabi2015maximizing} which considers the heterogeneous duration of communication time slots as a means to increase throughput. They assume a constant velocity of the mobile edge device and prove the NP-hardness of the optimization problem. The study relies on extensive simulations with various initial conditions to characterize the robustness of the online mixed integer linear programming approach. A three layer disaster rescue architecture combining UAV edge and fog computing for post-disaster rescue operations is proposed in \cite{sun2024joint} which address the problems of task offloading and resource allocation separately in a game-theoretic framework using convex optimization and evolutionary computation, but trajectory optimization is not considered. Hard deadlines for data sharing are the major constraint in \cite{samir2019uav}, which formulated the optimization problem to maximize the number of served devices while ensuring minimum data uploaded within the deadline.   

Edge computing is often used for computer vision task in traffic monitoring tasks \cite{wan2022edge} such as vehicle detection and tracking \cite{azfar2023incorporating}, speed and congestion classification \cite{liu2021smart}, cooperative perception \cite{zhang2019mobile}, etc.  An edge-enabled disaster rescue framework that uses computer vision to process crowdsourced photos on mobile edge devices in emergency vehicles was proposed in \cite{liu2019edge} incorporating an adaptive detection mechanism to handle unstable network conditions and demonstrated the system feasibility through a prototype implementation. Mobile edge computing using UAVs in a disaster stricken area was studied in \cite{shah2023mobile} focusing on network delay, quality of service, and network coverage. While these methods consider constraints related to limited battery sizes, quality of service, and communication urgency, the direct effects of the disaster scenario are not modeled or simulated.

Following the success of deep reinforcement learning in video games \cite{mnih2015human} and Go \cite{silver2017mastering}, the mobile edge computing research community has applied similar methods for decision-making for optimum resource allocation and path planning \cite{liu2020path} since the the different optimization problems are non-convex and the particular problems of task offloading and hover decision-making have been shown to be NP-hard by being analogous to the generalized assignment problem \cite{sun2022flexedge}. The optimization of UAV trajectories is studied in \cite{zhang2022deep} to serve IoT devices with communication and computing resources, aiming to minimize latency. Due to the complexity of the non-convex, nonlinear, and mixed discrete optimization problem, the authors propose two deep reinforcement learning algorithms that jointly optimize UAV path planning, user assignment, and resource allocation. Extending the problem to multiple UAVs, \cite{zhao2022multi} proposed multi-agent RL with Deep Deterministic Policy Gradient to learn the minimum energy consumption scheme for path planning and task offloading. A special consideration of task urgency appears in the reward function used by \cite{zhang2021task} for end-to-end deep RL to optimize UAV trajectory and task offloading. Resilience under communication disruption and physical obstacles is examined using RL \cite{tang2023disaster}, under a cooperative framework with edge agents also acting as network relays.

Although several studies have explored disaster scenarios as the context for UAV-assisted Mobile Edge Computing (MEC) for task offloading and communication, none explicitly address the direct impact of disasters on terrestrial edge devices. To the best of our knowledge, this work is the first to account for power and communication failures in edge devices and propose UAV assistance as a mitigation strategy. Using the hardware assumptions present in literature \cite{liu2020path}\cite{sun2023optimal} such as device types, tasks, and communication model \cite{chen2015efficient}, we introduce different configurations and extended the scope to include partial power supply, data expiry constraints, and identify the most critical devices to be serviced first to maximize the overall system lifetime. The objective is to develop a realistic strategy that enhances edge infrastructure resilience in emergency situations by incorporating practical constraints. Additionally, we simulate vehicle evacuations in a real-world setting, demonstrating how our approach extends the operational lifespan of edge devices while providing maintenance personnel with critical information on devices requiring attention.

\section{System Overview}

\begin{figure}[!htb]
    \centering
    \includegraphics[width=0.9\linewidth]{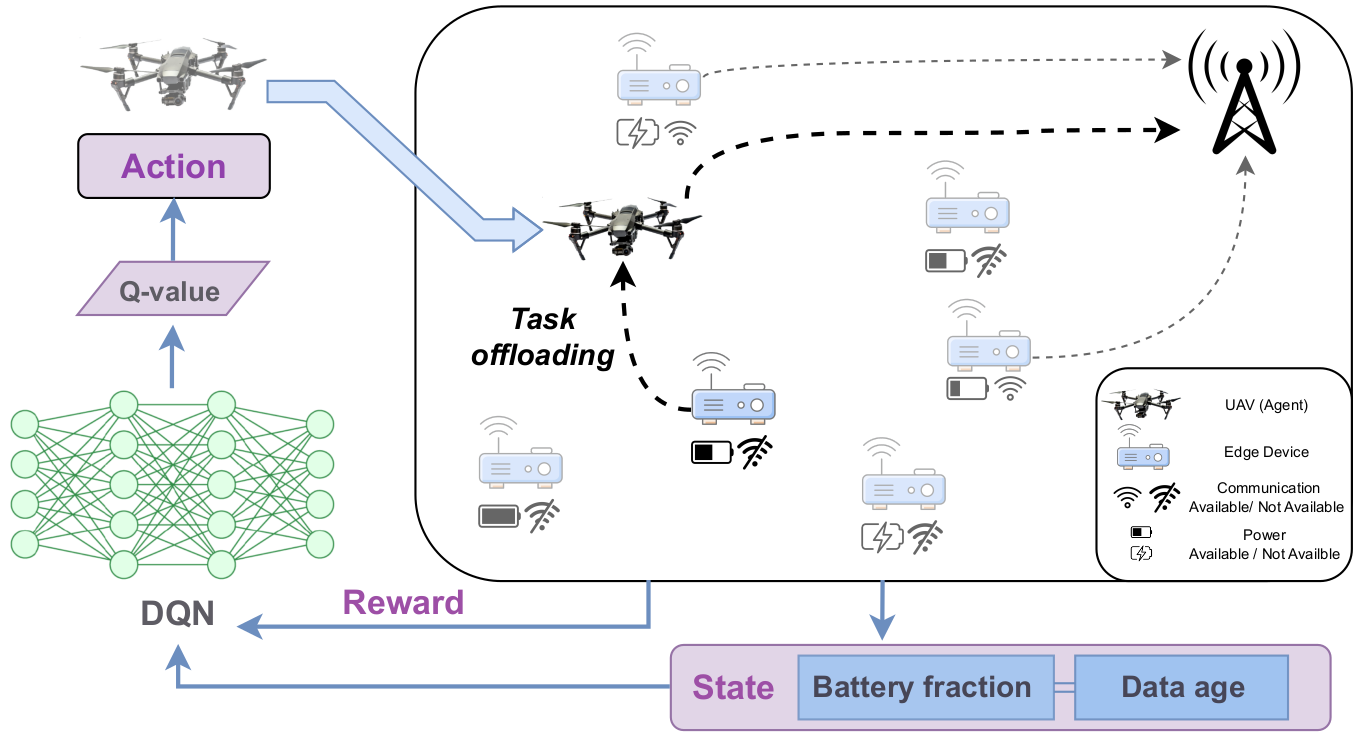}
    \caption{Proposed System overview}
    \label{fig:system}
\end{figure}

The system under consideration is composed of heterogeneous edge computing devices in a fixed rectangular region affected by a calamity that limits the power or communication to some or all edge devices. Figure~\ref{fig:system} shows the proposed system formulation. The edge devices perform video analytic tasks, and their power consumption rates for computation, communication, and standby are known in advance. The rate at which each device processes data locally is also known in advance, and is considered orders of magnitude slower than the UAV processing speed. The UAV is then deployed to this region and is able to move closer to any edge device to provide a communication channel and offloads the task data to execute the computation instead of the edge device. Depending on the scenario, an edge device may be deprived of power and thus unable to recharge, or lacking a communication channel, or both. 

\subsection{Problem Formulation}

The system is simulated with time slots based on the time taken for the UAV to travel to another node and the time taken for the edge devices to complete their task. Since the moving time of the UAV and local processing time of edge devices is not fixed, the time slot can have variable duration. The edge devices have limited battery, and the aim of the UAV control for task offloading is to maximize the battery life of all devices. If a device is visited, it consumes less power during that time slot because the UAV does its computation.
At the beginning of a time slot each edge device is assigned a volume of data randomly sampled from a bounded set to process, the UAV flies to a chosen device, hovers in place while offloading data and performing computation, then waits for the last device in the region to finish its local computation, which ends the time slot. The UAV is assumed to have a fixed speed and communication range. If multiple devices are present within communication range, they may all get served by the UAV in that time slot. In practice however, this is quite rare due to the random assignment of location and the limited communication range of edge devices. The goal of the system is to maximize the number of time slots of the UAV subject to some constraints. Formally, $x_{ij}^t$ is a binary variable that is 1 if the UAV moves from node $i$ to node $j$ at time slot $t$, and 0 otherwise. This information is formulated in statements (1a) and (1b):

\begin{subequations}
\begin{eqnarray}
&\max \ \sum_t \sum_i \sum_j x_{ij}^t \\
 \text{s.t.}  &x_{ij}^t  \in \{0, 1\} \quad \forall i,j,t \\
 &\sum_j x_{ij}^{t+1} = \sum_j x_{ji}^{t}  \quad \forall i,t\\ 
 &E_j(t+1) = E_j(t) - \delta_{Ej} \left[ \sum_{i} x_{ij}^t \ E_{\text{offload}}(j) - (1-\sum_{i} x_{ij}^t) \ E_{\text{process}}(j) \right] \quad \forall j, t \\
 &E_j(t+1) \geq 0 \quad \forall j, t \\
 &D_j(t+1) =  ( D_j(t) +  \delta_{Cj} ) (1 - \sum_{i} x_{ij}^t )  \quad \forall j, t \\
 &D_j(t+1) \leq D_{\text{max}} \quad \forall j, t 
\end{eqnarray}
\end{subequations}

Constraint (1c) ensures that the UAV destination node of the current time slot is the origin node of the next time slot. Equation (1d) indicates how the $j$th edge device loses battery ($E_j$) per time slot depending on whether it is the one being served by the UAV and consuming battery energy due to communication ($E_{\text{offload}}(j)$) or doing its own processing and consuming $E_{\text{process}}(j)$, where $\delta_{Ej}$ is a binary indicator that is 1 if the device has no power supply. Constraint (1e) requires that the battery level does not fall below zero. 

It is further assumed that due to the critical nature of the disaster scenario timely updates are needed from these devices as they are part of the intelligent infrastructure, and the data they provide can be valuable for rescue and recovery operations. Thus, (1f) says that if a device does not have communication capabilities ($\delta_{Cj} = 1$), and it is not visited by the UAV at time slot $t$, the data age $D(t)$ is incremented, but if it is visited then the new data age is zero. Then, (1g) enforces an upper limit to the data age for a device, $D_{\text{max}}$, and the expired data is considered as a failure of the system which is a constraint not usually considered in the task offloading and trajectory optimization literature. 

At the start of the simulation the edge devices are spawned with random architecture, e.g., Raspberry Pi, Jetson Nano, etc., random battery capacity chosen from predefined bounds, a randomly assigned computer vision task, e.g., Haar cascades, object detection with YOLOv3, etc., and placed in a random location uniformly chosen within the rectangular region. The UAV is initialized at the origin, which is one corner of the region. We then deviate from the literature by randomly assigning power and communication availability to edge devices to support the scenario analysis detailed in the next section. The distribution of every random parameter variable is uniform in its domain.

\subsection{Reinforcement Learning}
The logical next question to ask is: For a given configuration of the simulation described, what is the series of actions that the UAV can take such that the lifetime of the system is maximized? In other words, what is the best way to delay battery depletion and data expiry for all devices?
As this problem involves task offloading and minimizing energy consumption, and the added power and data expiry constraints do not make the problem any easier, in congruence with the existing literature, we use reinforcement learning (RL) to determine the best course of action for the UAV. In particular, deep Q-network (DQN) reinforcement learning is used with a multilayer perceptron neural network to learn the policy. 

In RL an agent learns to make decisions by interacting with an environment. The agent takes actions, receives feedback in the form of rewards, and aims to learn a policy that maximizes cumulative rewards over time. Q-learning is a model-free RL algorithm used to find the optimal action-selection policy for a given finite Markov decision process (MDP). It learns a function $Q(s, a)$, which represents the expected future rewards of taking action $a$ in state $s$, and aims to maximize this function from the direct sampling of the system outputs without estimating a model for the system itself. In the case of DQN, deep learning on a neural network is used to approximate the Q-function. The input to the neural network is the state, and the output is the predicted Q-value for each action. For training stability there are two neural networks: the Q-network and the target network. The Q-network is updated frequently with each iteration of training, while the target network is only periodically updated with the parameters of the Q-network. The target network is essentially a copy of the Q-network that is used for estimating Q-values and its purpose is to provide stable targets for the Q-values during training. Using a single neural network for both estimating Q-values and generating target Q-values can lead to irregularities during training in the form of divergence or oscillations. 

During every training iteration there is a reward associated with the action taken, which is feedback from the environment that indicates how good or bad an action was. Rewards can be positive, negative, or zero, depending on the task. Given an action ($a_t$) in the current state ($s_t$), the maximum Q-value achievable from the next state ($s_{t+1}$) is multiplied by a discount factor ($\gamma$) and added to the current reward ($r_t$) to give the target Q-value for the current state-action pair.

\begin{equation}  
Q_{\text {target }}\left(s_t, a_t\right)=r_t+\gamma \max _{a^{\prime}} Q\left(s_{t+1}, a^{\prime}\right)
\end{equation}

Thus, the Q-value represents the expected cumulative rewards of taking an action in a particular state and following the optimal policy thereafter. The Q-value of a state-action pair is predicted from a deep neural network which is trained using a loss function based on the difference of the predicted Q-value and the target Q-value.

\begin{equation}
    L(w) = (Q_w(s,a) - Q_{\text {target }}(s,a))^2
\end{equation}
where $L(w)$ is a loss function of the weights $w$ of the neural network which approximates $Q$ over the space of all state action pairs.

An optimal policy has the property that whatever the initial state and initial decision are, the remaining decisions must constitute an optimal policy given the state resulting from the first decision. However, selecting the action which results in the highest estimated Q-value during learning is not sufficient to discover the optimum policy, therefore techniques like epsilon-greedy exploration are commonly used, where the agent selects a random action with some probability epsilon ($\varepsilon$) to encourage exploration. In every iteration, with probability $\varepsilon$, the agent chooses to \textit{explore} by selecting an action uniformly at random from the action space or \textit{exploit} by selecting the action with highest expected Q-value otherwise. The value of $\varepsilon$ is decayed over time to gradually reduce the amount of exploration and increase the proportion of exploitation as the agent learns more about the environment. 
After the training is completed, the simulation uses the trained Q-network to choose the action with the highest predicted Q-value for any given state. In the testing phase, there are no random actions. This final simulation ends when one device fails, either due to a drained battery or expired data. 

\textbf{State: } A vector of length twice the number of devices, containing the remaining battery fraction and the age of uncommunicated data for each device. In the tested case this was 12 floating point numbers between 0 and 1 for the remaining battery fraction of each device, followed by 12 integers representing the age of the data in each device. At the first step of an episode the vector is initialized to $[1,1,1,1,1,1,1,1,1,1,1,1,0,0,0,0,0,0,0,0,0,0,0,0]$.

\textbf{Action: } An integer from 0 to 11 inclusive corresponding to the index of which device the UAV should visit in the next time slot.

\textbf{Reward: } The reward $r$ at time $t$ is calculated from $A_{t}$ the age of data on the device being visited at time $t$, $U_t$ the number of time slots the UAV has completed, $M_t$ the minimum battery fraction out of all the devices, and $O_t$ the age of the oldest data in the set. A number of different reward functions were explored and the results summarized in Table \ref{rewards}. A reward function can look like:
$$r_t =  \log(A_{t}) + \log(U_t) + \log(M_t) - \log(O_t)$$

\textbf{Q-network: } Fully connected neural network composed of two hidden layers with 64 neurons, each with ReLU activation. Larger problem size with more devices could require a larger network, but convergence to favorable performance was consistently observed with this network.

\section{Implementation}
The system is initialized with fixed random number seed for repeatability and multiple seeds were tried in each power and communication availability combination to obtain the average performance. The $x$ and $y$ coordinates of the devices are chosen uniformly at random on the grid, with random device type, random task type, and random maximum battery capacity, from the ranges shown in Table \ref{params}. 

In every time slot each device is assigned a random task volume for local processing or offloading to the UAV if it is visited. The UAV consumes power based on the traveled distance and time spent hovering. The local processing time for an edge device is found by dividing the task volume by the processing rate of the device for that task as shown in Table \ref{data}. Then the time is multiplied by the corresponding power from Table \ref{power} (plus standby power) to obtain the energy used for local processing in that time slot. The maximum processing time, and hence the maximum time slot duration, is manually capped at 10 minutes. If the device is able to communicate, the size of the data it transmits is assumed to be only 1 kilobyte since the output of computer vision object detection is often a simple list of bounding box or segmentation coordinates and labels in a small text file. If the device is not able to communicate, its data age is incremented by one.

If, instead, the device is visited by the UAV for task offloading, the edge device battery consumption depends on the data transmission for that time slot. The channel gain, $C$, is defined as $C=\beta_0 l^{-\theta}$, where $l$ is the distance between the edge device and UAV, $\beta_0$ is received power per unit distance, and $\theta$ is the path loss factor which models the attenuation of signal strength as it propagates through the wireless channel. The transmission rate, $R$, can then be calculated from Shannon's theorem \cite{shannon1949communication} as

\begin{equation}
R = W \log_2 \left( 1 + \frac{pC}{\sigma^2} \right)
\end{equation}

Where $W$ is the channel bandwidth, $p$ is the transmission power, and $\sigma^2$ is the channel noise power. Values for these parameters may be found in Table \ref{params} based on existing literature such as \cite{chen2015efficient}. In practice, these values should be calibrated to the specific scenario to better estimate the transmission rate. The task volume is then divided by $R$ to find the transmission time.

Deep Q-learning is performed in a Gym environment in Python using the stable baselines reinforcement learning library \cite{raffin2021stable}. The environment contains the system model described above, and every step of the simulation starts with choosing an action, i.e., which device to visit, then that device offloads data while the rest process their tasks locally. While devices wait for the last one to finish processing, they consume only standby power. An episode of simulation consists of a series of time slots and ends when any device runs out of battery or the age of data in any device reaches 10 time slots. The exploration factor is linearly reduced from 1 to 0.05 over the course of the first 35\% of the training iterations and then remains at that value. The parameters of the deep Q-learning are shown in Table \ref{rlparams}.

\begin{table}[!htpb]
\centering
\caption{Simulation Parameters}
\begin{tabular}{lc}
\hline
Parameter                        & Value                                                               \\ \hline
UAV speed                        & 5 m/s                                                               \\ 
UAV flying power   consumption   & 150 W                                                               \\ 
UAV hovering power   consumption & 80 W                                                                \\ 
UAV signal range                 & 65 m                                                                \\ 
UAV altitude                     & 10 m                                                                \\
Device types                     & \begin{tabular}[c]{@{}l@{}}Raspberry Pi 4B, Raspberry Pi 3B, \\ Firefly, Jetson   Nano, NanoPC-T4\end{tabular} \\ 
Standby power   (respectively)   & [2.65, 1.69, 4.87, 2.82, 1.88] Watts                            \\ 
Number of edge   devices         & 12                                                                  \\ 
Transfer bandwidth, $W$            & 20 MHz                                                              \\ 
Channel noise power $\sigma$     & -100 dBm                                                            \\ 
Received power per   meter, $\beta_0$   & -50 dBm                                                             \\ 
Path loss factor, $\theta$              & 4                                                                   \\ 
Transmission power, $p$            & 0.1 W                                                               \\ 
Device battery capacity range         & 50 KJ – 80 KJ                                                       \\ 
UAV battery capacity             & 1600 KJ                                                             \\ 
Task volume range                & 2 MB – 4 MB                                                         \\ 
Data age limit                   & 10 time slots                                                       \\ \hline
\end{tabular}
\label{params}
\end{table}

The state vector consists of the battery fraction of each device and the age of the data in each device. At the end of the step a reward is calculated by calling one of the several reward functions listed in Table \ref{rewards}. The most consistently favorable reward was found to be the sum of the number of time slots elapsed and the log of the ratio of the age of data in the chosen device to the oldest data in any device. The intuition behind the chosen rewards is that the factors added to the reward are desired to be increased (e.g., UAV time steps, minimum battery), while the factors subtracted should be decreased(e.g., oldest data age), and the deep neural network will learn this over hundreds of thousands of iterations and take the action that is most likely to maximize the episode length. 

After the training the simulation is run deterministically using the trained network to choose the action, and the device that fails first is identified. In a practical scenario, this would inform the responders to provide maintenance to this device first, as it is the critical point of failure in that particular configuration. Of course, this assumes that all devices are of equal importance in terms of the tasks they perform. In case the failing device is not critical to the system, it may be removed, and the reinforcement learning can be run again to determine the critical device that should receive priority attention.

\begin{table}[!htpb]
\centering
\caption{ Power for each device by task, in Watts \cite{sun2023optimal} }
\begin{tabular}{lccccc}
\hline
\multicolumn{1}{l}{} & \multicolumn{1}{l}{Raspberry Pi 4B} & \multicolumn{1}{l}{Raspberry Pi 3B} & \multicolumn{1}{l}{Firefly} & \multicolumn{1}{l}{Jetson Nano} & \multicolumn{1}{l}{NanoPC-T4} \\ \hline
HAAR                   & 2.165                                & 1.287                                & 2.352                        & 1.36                             & 3.391                          \\ 
MMOD                   & 1.335                                & 1.621                                & 1.124                        & 0.92                             & 1.582                          \\ 
DNN                    & 3.234                                & 1.9                                  & 2.972                        & 2.421                            & 2.595                          \\ 
Dlib                   & 1.91                                 & 4.38                                 & 2.54                         & 1.01                             & 5.06                           \\ 
YOLOv3                 & 3.268                                & 1.925                                & 3.07                         & 1.35                             & 2.874                          \\ \hline
\end{tabular}\label{power}
\end{table}

\begin{table}[!htb]
\centering
\caption{Data processing rate for each device and task in Bytes per second \cite{sun2023optimal}  }
\begin{tabular}{lccccc}
\hline
       & Raspberry Pi 4B & Raspberry   Pi 3B & Firefly  & Jetson   Nano & NanoPC-T4 \\ \hline
HAAR   & 74536.25        & 2758.14           & 4734.16  & 23867.61      & 1854.59   \\ 
MMOD   & 12318.36        & 1114.03           & 1183.72  & 5277.71       & 949.66    \\ 
DNN    & 71731.84        & 3767.51           & 949.36   & 40294.21      & 973.1     \\ 
Dlib   & 65088.52        & 87894.05          & 13020.39 & 65264.07      & 6957.73   \\ 
YOLOv3 & 69985.94        & 7066.79           & 2733.97  & 22230.58      & 8298.13   \\ \hline
\end{tabular} \label{data}
\end{table}

\begin{table}[!htb]
\centering
\caption{Reinforcement learning parameters}
\begin{tabular}{lc}
\hline
Parameter                & Value     \\ \hline
State vector size        & 24        \\
Action space             & [0,11]        \\
Batch size               & 16        \\
Gamma                    & 0.98      \\
Learning rate            & 0.0071    \\
Learning iterations      & 1,000,000 \\
Exploration fraction     & 0.35      \\
Minimum exploration rate & 0.05      \\ \hline \\
\end{tabular} \label{rlparams}
\end{table}

\section{Results and Case Study}

\begin{figure}[!htpb]
     \centering
     \includegraphics[width=0.75\linewidth]{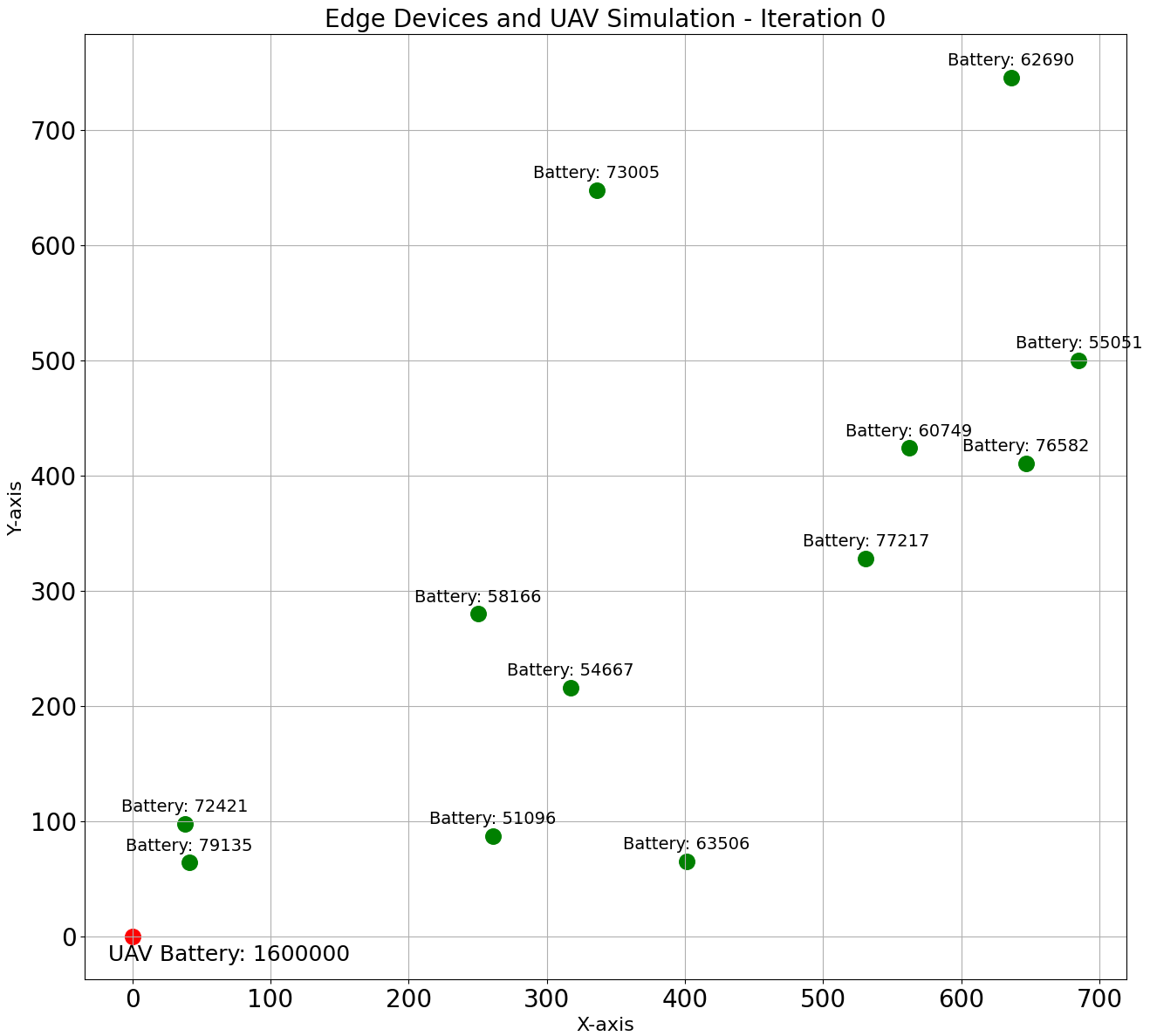}
     \caption{Example layout of edge devices in the region with the start position of the UAV at the origin. Distance is in meters, battery values are in Joules.}
     \label{fig:devices}
 \end{figure}

The experiments were performed on Intel Core i9 13900-K CPU with 32 GB RAM and NVIDIA GeForce RTX 4090 GPU using Python 3.11.5, and each training run took approximately 11 minutes. The size of the region chosen directly affects the episode length as the battery of the UAV is also limited, albeit much greater than those of edge devices. After some trial and error, 800m by 800m was chosen so that the UAV range was not the limiting factor. One initial layout is depicted in Figure \ref{fig:devices}, showing 12 edge devices with initial battery capacity and the UAV at its starting position.

\begin{figure*}[!htpb]
    \centering
    \begin{subfigure}[t]{0.85\textwidth}
        \centering
     \includegraphics[width=0.9\linewidth]{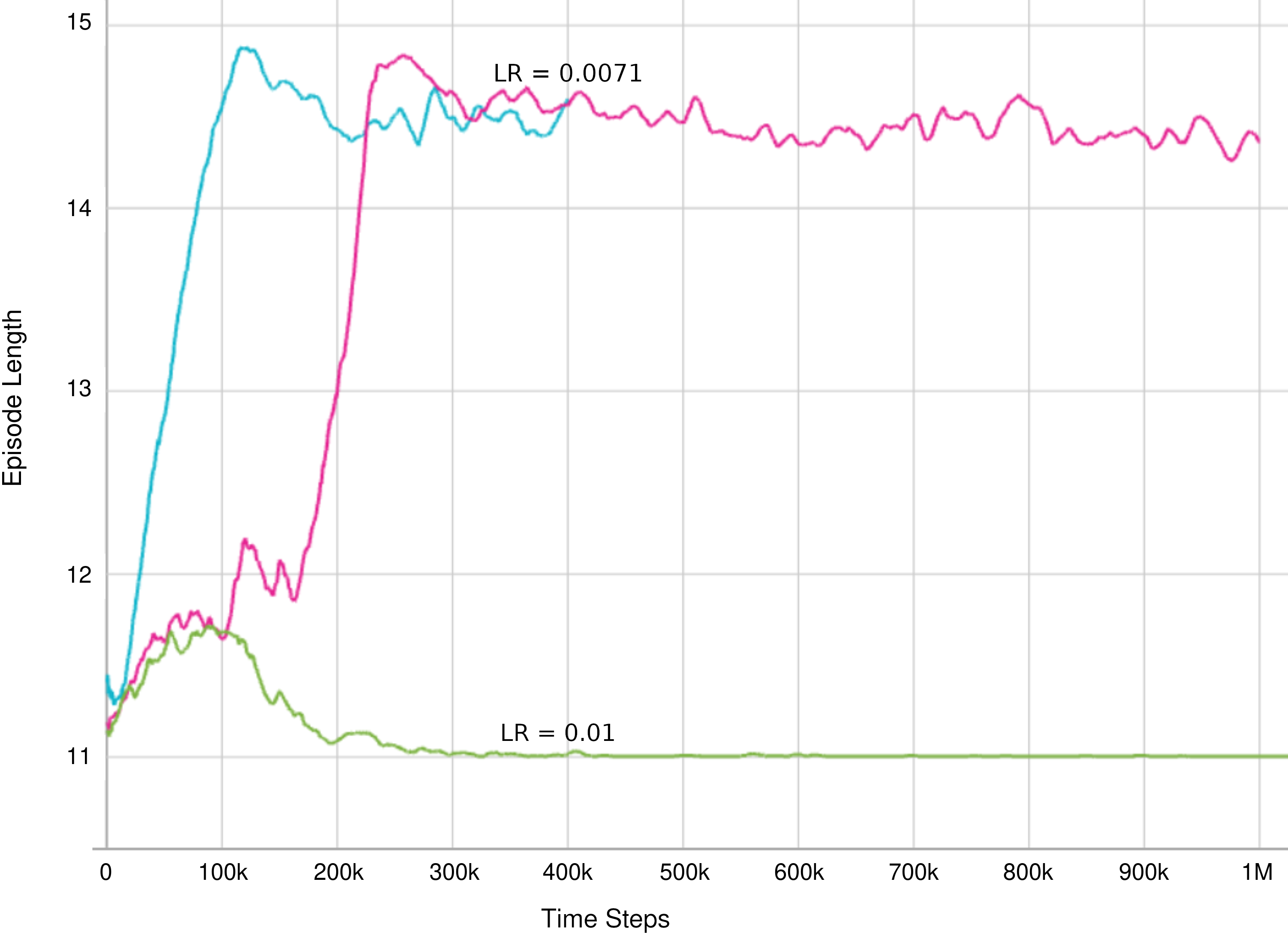}
     \caption{ Two learning rates with the same initial conditions. The cyan curve is run for 400,000 iterations, showing that in some scenarios shorter training times can achieve the same performance.}
     \label{fig:learnrate}
    \end{subfigure}%
\\
    \begin{subfigure}[t]{0.85\textwidth}
        \centering
     \includegraphics[width=0.9\linewidth]{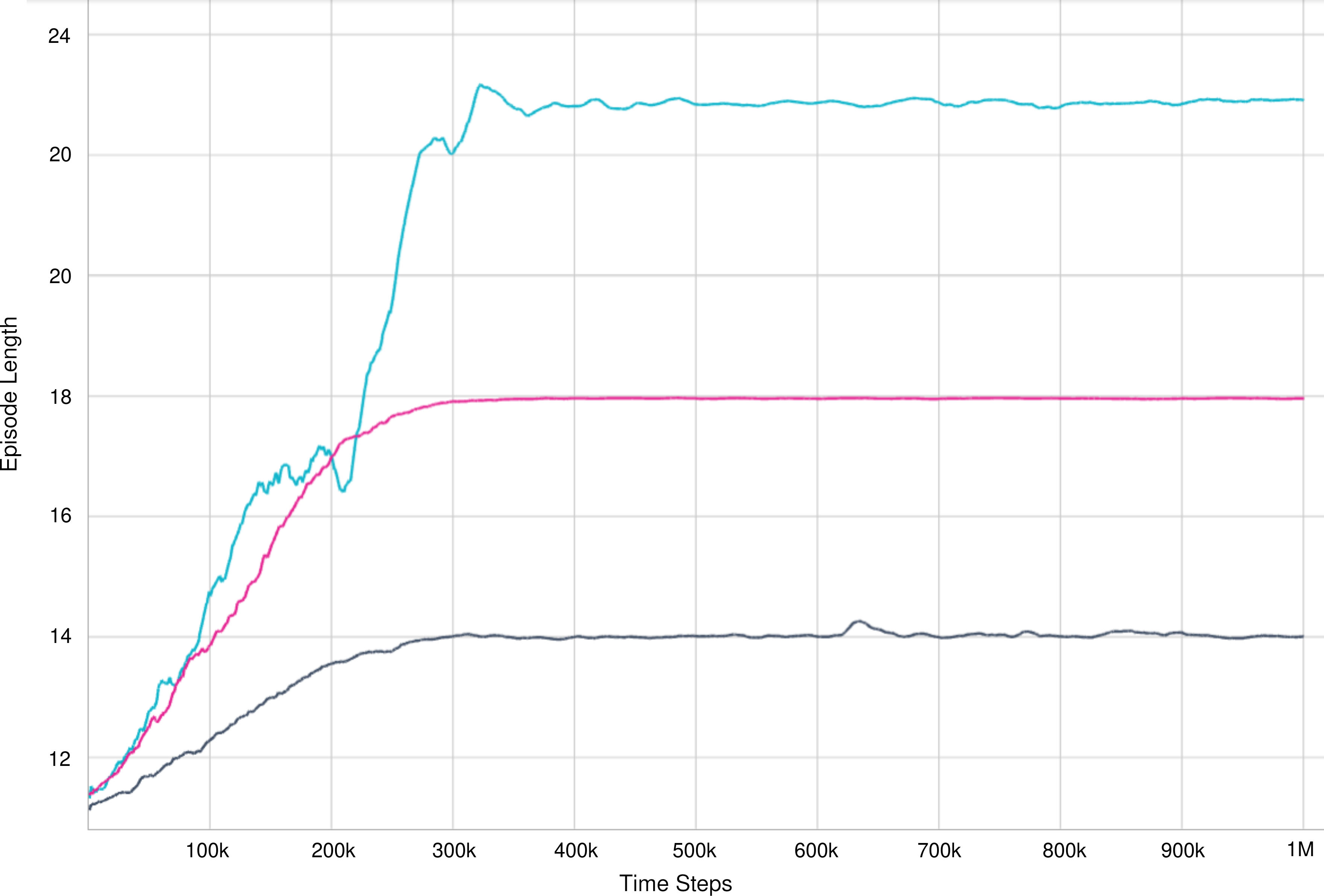}
     \caption{Three different random initial conditions for 8 devices with power supply and 8 devices with communications enabled.}
     \label{fig:initialCondition}
    \end{subfigure}
    \caption{Reinforcement learning training showing the increase in average episode length as the DQN learns the best policy.}
\end{figure*}

Some hyperparameter tuning was required for successful execution. The learning rate for the deep Q-network was decreased from 0.01 to around 0.007 which provided stable convergence to the maximum attainable episode length in every scenario. The behavior of the reinforcement learning under the two learning rates is shown in Figure \ref{fig:learnrate}. We note that the model converges to the best strategy in a lower number of iterations in some scenarios, but the value of 1,000,000 iterations was chosen across the board for all experiments for consistency as it always converges in practice. 

The maximum life of the system with full power and communication is infinite if we only consider the edge devices, but is limited in simulation by the battery life of the UAV. In practice, it can be assumed that the UAV will return to base and another one may be able to take its place. Table \ref{result1} shows the average lifetime (in time slots) of each combination of power and communication availability to the edge computing nodes. As a baseline, if no device has power supply or communication, the episode lasts for 11 time slots (since the data expires).

\begin{table}[!htpb]
\centering
\caption{\vspace{0.2cm}Average episode length for each scenario of power and communication availability}
\begin{tabular}{c|ccccc}
\hline
\diagbox[]{Power}{Comms} & 12   & 10   & 8    & 6    & 4    \\ \hline
12                       & 33   & 33   & 29   & 29   & 29   \\ 
10                       & 25   & 24.7 & 18.7 & 19   & 17.7 \\ 
8                        & 21.6 & 19.3 & 18.3 & 18   & 16.7 \\ 
6                        & 18.6 & 18   & 18   & 17.3 & 17   \\ 
4                        & 17   & 17   & 17.3 & 18   & 17.3 \\ \hline
\end{tabular} \label{result1}
\end{table}

Figure \ref{fig:initialCondition} shows the episode length versus iteration of reinforcement learning using three different random seeds. Note the significant effect of initial conditions such as device type, task type, device location, and which devices are chosen to be without power and/or communication. In all three curves, there are 4 devices lacking power and 4 devices lacking communication, and there is no limitation on these being the same or different devices. For each case there is a different maximum episode length possible under the UAV task offloading and communication scheme.

Multiple reward functions were tested in three scenarios of power and connection disruptions under three different RL methods with the results summarized in Table \ref{rewards}. DQN was observed to consistently outperform other methods, while the final reward listed in the table performs the best in combination with DQN across all scenarios. 

\begin{table}[!htpb]
\caption{\vspace{0.2cm}Maximum episode length for three RL methods in three configurations with various reward functions}
\label{rewards}
\begin{tabular}{l|lll|lll|lll}
\hline
\multicolumn{1}{c|}{Configuration}           & \multicolumn{3}{c|}{Comm 6, power 6} & \multicolumn{3}{c|}{Comm 8, power 8} & \multicolumn{3}{c}{Comm 8, power 10} \\ 
\multicolumn{1}{c|}{RL Method}              & DQN        & PPO        & A2C       & DQN        & PPO        & A2C       & DQN        & PPO        & A2C        \\ \hline \hline
\textbf{Reward}                &            &            &           &            &            &           &            &            &            \\
Data age of current device ($A$) & 16         & 11         & 11        & 19         & 19         & 12        & 18         & 11         & 11         \\
UAV Timesteps ($U$)              & 15         & 11         & 11        & 17         & 11         & 11        & 13         & 11         & 11         \\
Minimum battery ($M$)            & 12         & 11         & 11        & 11         & 13         & 11        & 23         & 11         & 16         \\
$-$ Oldest data age ($O$)          & 11         & 11         & 11        & 14         & 15         & 11        & 16         & 11         & 11         \\
$A+U+M-O$                        & 17         & 11         & 14        & 16         & 11         & 11        & 16         & 11         & 11         \\
log($A$)                         & 13         & 15         & 11        & 23         & 15         & 18        & 11         & 14         & 11         \\
log($U$)                         & 11         & 15         & 11        & 19         & 11         & 11        & 11         & 14         & 11         \\
log($M$)                         & 11         & 11         & 11        & 11         & 11         & 11        & 11         & 11         & 11         \\
$-$log($O$)                         & 11         & 11         & 11        & 18         & 11         & 11        & 28         & 14         & 11         \\
log($A$)+log($U$)+log($M$)-log($O$)    & 17         & 12         & 11        & 18         & 11         & 11        & 14         & 22         & 11   \\ 
$U$ + log$\left(\dfrac{A}{O}\right)$ & \textbf{18} & 15 & 11 & \textbf{28} & 18 & 12 & \textbf{30} & 21 & 12 \\ \hline     
\end{tabular}
\end{table}

\subsection{Prioritizing the Evacuation Route}
To evaluate the system's performance in a realistic disaster scenario, we designed an evacuation simulation for a small rural town in upstate New York called Round Lake. The simulation involved vehicle flows, defined using the SUMO (Simulation of Urban MObility) tool \cite{krajzewicz2002sumo}, from residential areas towards the main Adirondack highway. In this scenario there are some devices that are absolutely critical to the evacuation as they provide road usage information of high traffic areas, while other devices in areas that have already been evacuated may no longer need to be kept operational. 

\begin{figure*}[!htpb]
    \centering
    \begin{subfigure}[t]{0.85\textwidth}
        \centering
    \includegraphics[width=0.72\linewidth]{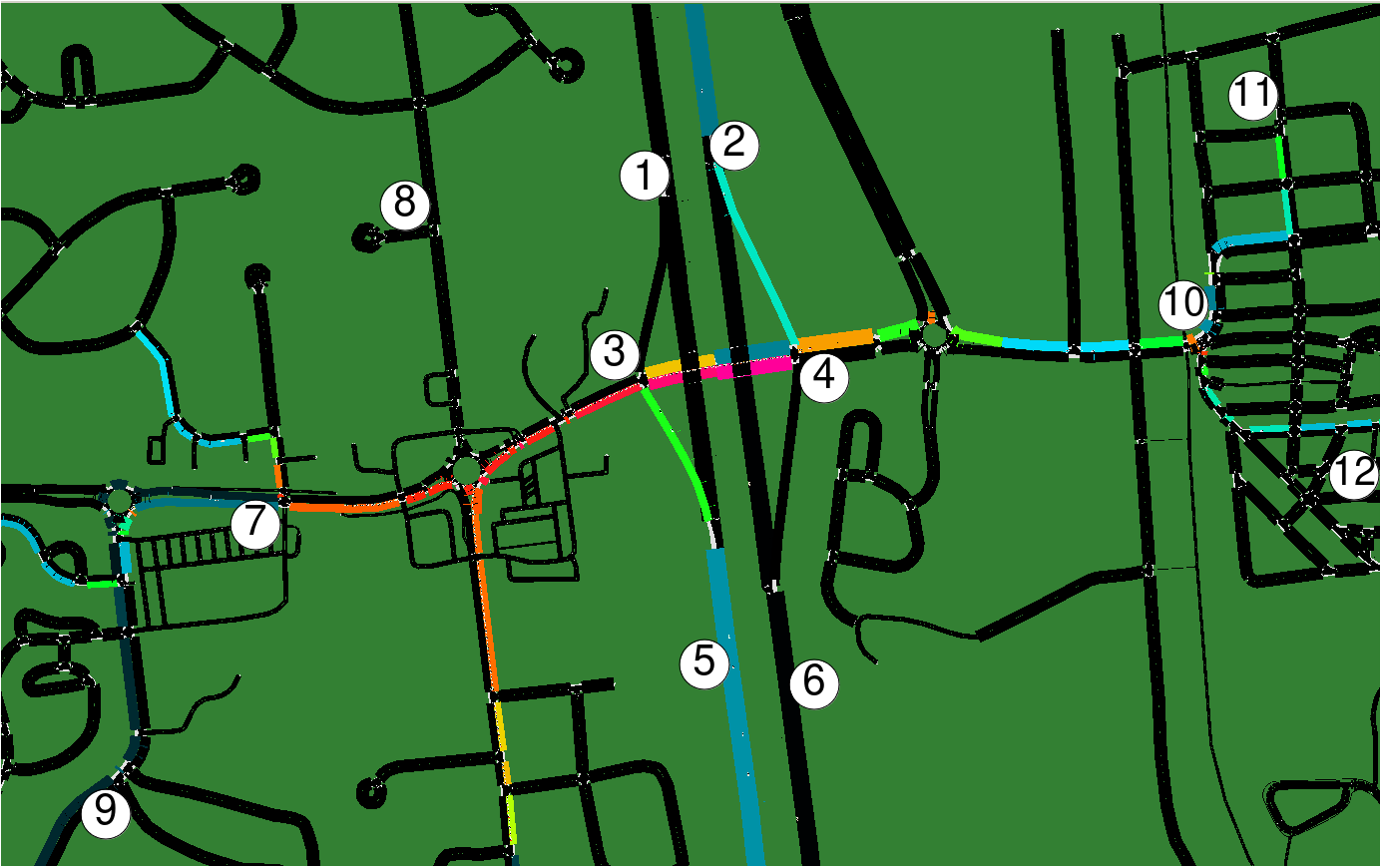}
    \caption{Rural road network}
        \label{fig:sumorural}
    \end{subfigure}%
    
    \begin{subfigure}[t]{0.85\textwidth}
        \centering
    \includegraphics[width=0.75\linewidth]{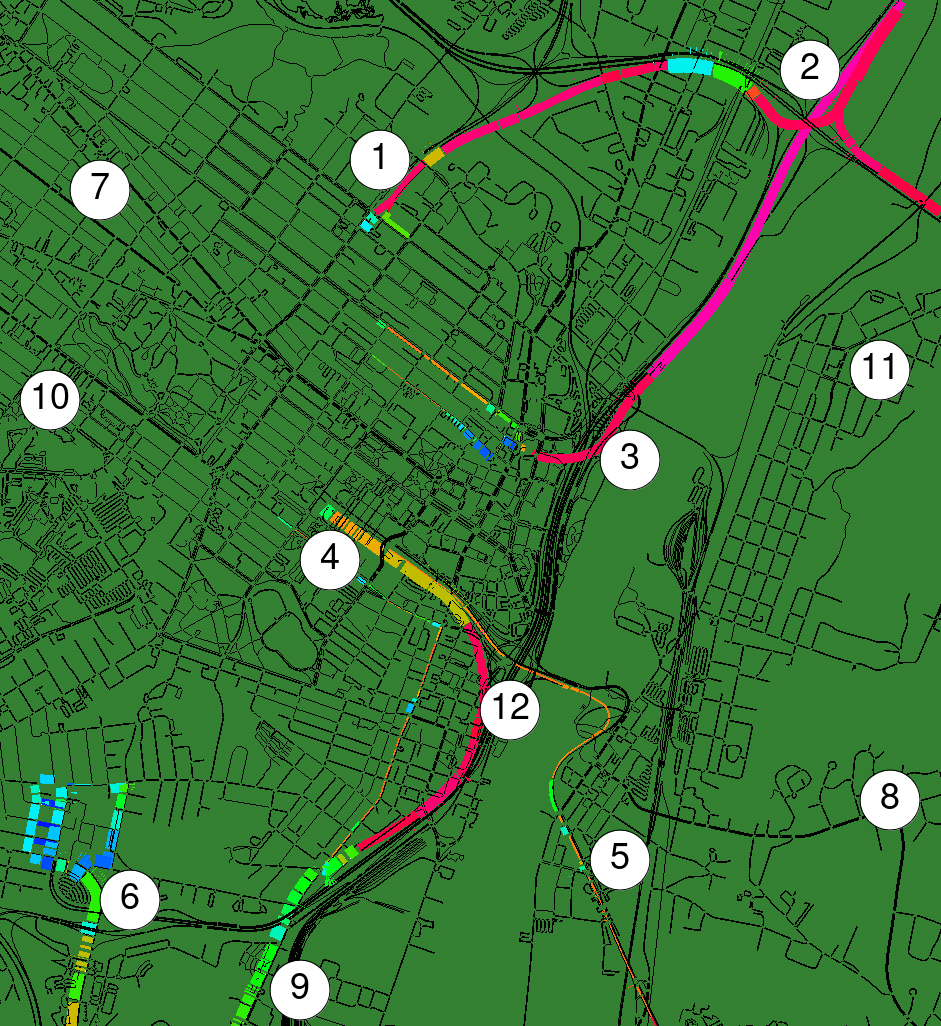}
    \caption{Urban road network}  
    \label{fig:sumourban}
    \end{subfigure}
    \caption{Road network with numbered edge devices. Traffic density on the roads is color coded in order of increasing density: {\color{blue}blue}, {\color{cyan}cyan}, {\color{yellow}yellow}, {\color{green}green}, {\color{orange}orange}, {\color{red}red}, {\color{magenta}magenta}. }
\end{figure*}

Therefore, edge devices were strategically placed at various junctions over the simulated road network, with some located near the predefined evacuation routes as shown in Figure \ref{fig:sumorural}. The SUMO simulation was executed to determine the traffic densities on these roads. The edge devices utilize computer vision to detect vehicles and vulnerable road users, thereby providing crucial information to emergency services. The UAV-edge RL system was programmed to receive additional rewards if an edge device was situated near a high-density road, in proportion to the observed average density. A similar experiment was also performed on the downtown Albany road network at a larger scale as shown in Figure \ref{fig:sumourban}. This also contains major corridors of evacuation while the rest of the traffic is considered to be negligible or low priority.

\begin{table}[!htpb]
\centering
\caption{\vspace{0.2cm}Failure rate of the devices deployed in the evacuation zone}
\begin{tabular}{ccccccccccccc}
\hline
 Device  & 1 & 2 & 3 & 4 & 5 & 6 & 7 & 8 & 9 & 10 & 11 & 12  \\ \hline
Rural    & 2 & 0 & 0 & 0 & 0 & 3 & 0 & 0 & 6 & 0  & 8  & 11   \\ 
Urban    & 0 & 0 & 0 & 3 & 0 & 1 & 6 & 10 & 5 & 3  & 2  & 0   \\  
\hline
\end{tabular} \label{table:failures}
\end{table}

% \begin{figure}
%     \centering
%     \includegraphics[width=0.85\linewidth]{failure.png}
%     \caption{Failure rate of the devices deployed in the evacuation zone}
%     \label{fig:failures}
% \end{figure}

The UAV was again tasked with providing computational offloading and communication support to these edge devices to support evacuation. For this scenario, 4 devices did not have access to power, while 6 devices did not have access to communication. Multiple tests were conducted with different random configurations for power and communication availability in each test to observe the trend of device failure. Reinforcement learning was run for 300,000 iterations for every test and a total of 30 tests were performed with different random seeds. 
Statistically, it was found that the devices that were allowed to fail most frequently were those not located on high-traffic roads despite the random arrangement of power and connectivity issues. 

The failure frequency is shown in Table \ref{table:failures}. For the rural scenario devices 2, 3, 4, 5, 7, and 10 are near high traffic zones and those were not the first to fail in any of the tests. There were a couple of anomalies: device 8 is not close to any traffic yet it was not observed to fail first in any test, while device 11 is close to low density but suffered a high failure rate. A similar trend was observed in the urban scenario where devices 1, 2, 3, and 12 are in the high density paths and they are prevented from failure. Again there are anomalies as devices 6 and 9 are in higher density regions than device 10 and 11 but still observe some failures. 

These results indicate that the UAV successfully learns to prioritize devices close to high traffic flow despite variations in power and communication constraints. However, there is room for further exploration especially in the design of the reward function, and the inclusion of a larger traffic network with granular traffic data for a more realistic simulation.

\section{Conclusion}

Intelligent edge computing networks are fast becoming the future architecture of transportation infrastructure that will be essential in disaster or emergency situations. In this paper we explore the aftermath of a disaster that disables power and communication access to some edge infrastructure devices with heterogeneity. The concept of a mobile edge node in the form of a UAV is used to provide computation offloading and communication relay to the affected edge nodes. The problem of finding the best course of action for a given situation is formulated, and addressed through deep Q-network reinforcement learning. The two main constraints are decreasing battery charge and the increasing age of uncommunicated data. Experiments are performed for multiple combinations of power and communication availability with a variety of reward functions and RL methods. While the reduction of the life of the network is expected with decreasing power and connectivity, the important outcome of each experiment is to identify the device that is expected to fail first, thus informing responders and operators of which devices to prioritize for maintenance. The system was also tested in conjunction with evacuating traffic in two settings, rural and urban, where higher priority was given to devices located near high flow roadways.

Future work will explore the deployment of multiple UAVs to further enhance the resilience and efficiency of the proposed system. By integrating multiple UAVs, we aim to improve the coverage and redundancy of computational offloading and communication relays, particularly in large or densely populated disaster zones. Additionally, we will investigate the feasibility of ad-hoc mesh networks, allowing UAVs and edge devices to dynamically form communication links without relying on fixed infrastructure. This could significantly enhance the robustness of the network during large-scale emergencies. We also plan to incorporate cost and logistical considerations into our optimization models, factoring in the complexity and expenses associated with deploying personnel for device maintenance and repair. By combining these elements, our goal is to develop a comprehensive, practical solution that optimizes both operational effectiveness and cost-efficiency in real-world disaster response scenarios. Further, we will expand our simulations to include a more extensive traffic network with granular data, enabling a more detailed assessment of the system's impact on evacuation and rescue operations. \\

%%
%% The acknowledgments section is defined using the "acks" environment
%% (and NOT an unnumbered section). This ensures the proper
%% identification of the section in the article metadata, and the
%% consistent spelling of the heading.

% \begin{acks}
% To Robert, for the bagels and explaining CMYK and color spaces.
% \end{acks}

%%
%% The next two lines define the bibliography style to be used, and
%% the bibliography file.
\bibliographystyle{ACM-Reference-Format}
\bibliography{sample-base}

%%
%% If your work has an appendix, this is the place to put it.
% \appendix

% \section{Research Methods}

% \subsection{Part One}

% Lorem ipsum dolor sit amet, consectetur adipiscing elit. Morbi
% malesuada, quam in pulvinar varius, metus nunc fermentum urna, id
% sollicitudin purus odio sit amet enim. Aliquam ullamcorper eu ipsum
% vel mollis. Curabitur quis dictum nisl. Phasellus vel semper risus, et
% lacinia dolor. Integer ultricies commodo sem nec semper.

% \subsection{Part Two}

% Etiam commodo feugiat nisl pulvinar pellentesque. Etiam auctor sodales
% ligula, non varius nibh pulvinar semper. Suspendisse nec lectus non
% ipsum convallis congue hendrerit vitae sapien. Donec at laoreet
% eros. Vivamus non purus placerat, scelerisque diam eu, cursus
% ante. Etiam aliquam tortor auctor efficitur mattis.

% \section{Online Resources}

% Nam id fermentum dui. Suspendisse sagittis tortor a nulla mollis, in
% pulvinar ex pretium. Sed interdum orci quis metus euismod, et sagittis
% enim maximus. Vestibulum gravida massa ut felis suscipit
% congue. Quisque mattis elit a risus ultrices commodo venenatis eget
% dui. Etiam sagittis eleifend elementum.

% Nam interdum magna at lectus dignissim, ac dignissim lorem
% rhoncus. Maecenas eu arcu ac neque placerat aliquam. Nunc pulvinar
% massa et mattis lacinia.

\end{document}